\documentstyle[aaspp4,11pt,flushrt,epsf]{article}
\begin{document}
\title{An Analytic Expression for the Growth Function in a 
Flat Universe with a Cosmological Constant}
\author{Daniel J.\ Eisenstein}
\affil{Institute for Advanced Study, Princeton, NJ 08540}

\begin{abstract}
An analytic expression is given for the growth function for
linear perturbations in a low-density universe made flat by a
cosmological constant.  The result involves elliptic integrals but is
otherwise straight-forward.
\end{abstract}

\keywords{cosmology: theory -- large-scale structure of the universe}

In linear cosmological perturbation theory, the evolution of perturbations
in the absence of pressure
is reduced to the superposition of two modes with fixed time dependence and
arbitrary spatial dependence (\cite{Pee80}\ 1980).  The behavior of these modes 
in time is a function of the background cosmological parameters.  The
functions of time that describe this behavior are called the growth
functions and denoted $D_1$ and $D_2$.

As shown by \cite{Hea77}\ (1977), the growth functions in cosmologies where
pressure fluctuations are negligible can be written as 
\begin{eqnarray}\label{eq:Done}
D_1(a) &\propto& H(a)\int^a_0 {da'\over a'^3 H(a')^3},\\
D_2(a) &\propto& H(a),\\
H(a) &=& \sqrt{\Omega_0 a^{-3} + \Omega_R a^{-2} + \Omega_\Lambda}.
\label{eq:H}
\end{eqnarray}
Here, $a$ is the expansion scale factor of the universe, chosen so that
$a=1$ today; this will be our time coordinate.  The redshift $z$ is
then $a^{-1}-1$.  $\Omega_0$ is the
density of non-relativistic matter in the universe today in units of
the critical density.  $\Omega_\Lambda = \Lambda/3H_0^2$ is the
cosmological constant $\Lambda$ relative to the
present-day Hubble constant $H_0$.  $\Omega_R$ is the curvature term
and is equal to $1-\Omega_0-\Omega_\Lambda$.  We assume that relativistic 
matter is a negligible contributor to the density of the universe.
Finally, $H(a)$ is proportional to the Hubble constant at epoch $a$,
scaled so that $H=1$ today.  We are interested in the behavior of
$D_1(a)$, as this is the mode whose amplitude grows in time.  However,
the overall normalization of $D_1$ is merely a convention.

The analytic form for $D_1$ in the $\Lambda=0$ case
(\cite{Wei72}\ 1972; \cite{Gro75}\ 1975; \cite{Edw76}\ 1976) is
widely known, but no analytic solution for the case of $\Lambda\ne0$
has been presented in the literature.  Instead, workers integrate
equation (\ref{eq:Done}) numerically or use the approximations
given by \cite{Lah91}\ (1991) or \cite{Car92}\ (1992).  
However, the integral can be done in
terms of elliptic integrals, which of course are implemented as special
functions in many numerical packages.  Here, we restrict ourselves to
the case of a flat, low-density universe, i.e.\ $\Omega_R=0$,
$\Omega_0<1$, $\Omega_\Lambda=1-\Omega_0$.  This is the case commonly
used in practice and the solution can be stated rather cleanly.

We adopt a normalization for $D_1$ as
\begin{equation}\label{eq:norm}
D_1(a) = {5\Omega_0\over2} H(a)\int^a_0 {da'\over a'^3 H(a')^3}.
\end{equation}
This is chosen so that $D_1(a)\rightarrow a$ as $a\rightarrow0$.
Defining $v = a^{-1}\sqrt[3]{\Omega_0/(1-\Omega_0)}$ and
manipulating the integral yields
\begin{eqnarray}
D_1(a) &=& {5\over2}av\sqrt{1+v^3}\int^\infty_v{u\,du\over(1+u^3)^{3/2}}\\
&=& {5\over2}av\sqrt{1+v^3}
\lim_{v_\infty\rightarrow\infty} \lim_{b\rightarrow1}
(-2){d\over db}\int^{v_\infty}_v{u\,du\over\sqrt{b+u^3}}\\
&=& 5av\sqrt{1+v^3}\lim_{v_\infty\rightarrow\infty} \left[
-{1\over6}\int^{v_\infty}_v{u\,du\over\sqrt{1+u^3}}
+{1\over3}{v_\infty^2\over\sqrt{1+v_\infty^3}}-{1\over3}{v^2\over\sqrt{1+v^3}}
\right].
\end{eqnarray}
Doing this last integral (e.g.\ using 3.139.8 in \cite{Gra94}\ 1994) 
yields the final answer
\begin{equation}\label{eq:sol}
D_1(a) = a \times d_1\left({1\over a}\sqrt[3]{\Omega_0\over1-\Omega_0}\right),
\end{equation}
where
\begin{eqnarray}\label{eq:done}
d_1(v) &=& {5\over3}v\left\{
\sqrt[4]{3}\sqrt{1+v^3}\left[E(\beta,\sin75^\circ)-
{1\over3+\sqrt{3}}F(\beta,\sin75^\circ)\right]+
{1-(\sqrt{3}+1)v^2 \over v+1+\sqrt{3}}\right\},\\
\beta&=&{\rm Arccos}\,{v+1-\sqrt{3}\over v+1+\sqrt{3}}.
\end{eqnarray}
Here, $F$ and $E$ are incomplete elliptic integrals of the first and
second kind, using the definitions
$F(\phi,k)=\int^\phi_0\,(1-k^2\sin^2\theta)^{-1/2}d\theta$
and $E(\phi,k)=\int^\phi_0\,(1-k^2\sin^2\theta)^{1/2}d\theta$; 
note that conventions vary on whether $k$ is squared
or not.  Also note that $\beta$ can exceed $\pi/2$; some numerical
implementations of the elliptic integrals do not handle this 
properly.  In this case one can use the identity 
$F(\phi,k)=2F(\pi/2,k)-F(\pi-\phi,k)$ and likewise for $E$ to compute
for $\beta>\pi/2$.

As $a\rightarrow0$, $v\rightarrow+\infty$ and $d_1\rightarrow1-(2/11)v^{-3}
+ (16/187)v^{-6} + O(v^{-9})$.  This is not obvious from equation
(\ref{eq:done}); the terms of order $v^2$ and $v$ cancel.  This suggests
that for numerical evaluation for $v\gg1$, one should be watchful for
loss of precision.  Alternatively, for $v\gtrsim5$ 
one can use the large $v$ expansion
given above.  As $a\rightarrow\infty$, $v\rightarrow0$ and
$D_1 = 1.4373\sqrt[3]{\Omega_0/(1-\Omega_0)}$.

The approximation of \cite{Car92}\ (1992, adapted from 
\cite{Lah91}\ 1991) that
\begin{equation}
D_1(1) = {5\Omega_0\over2}\left[\Omega_0^{4/7}-\Omega_\Lambda+
(1+\Omega_0/2)(1+\Omega_\Lambda/70)\right]^{-1}
\end{equation}
can be compared to $d_1(v)$ using $\Omega_\Lambda=1-\Omega_0=(v^3+1)^{-1}$.
The fit has fractional errors better than 2\% over the range $z\ge0$ 
and $\Omega_0>0.1$ (i.e.\ $v>0.48$) but deviates for lower $\Omega_0$.  
A plot of $d_1(v)$ and the \cite{Car92}\ (1992) fit is shown in
Figure 1.  The fit has similar accuracy for the $\Lambda=0$ case even
to much lower $\Omega_0$.

For completeness, the solution for the expanding epoch of a flat universe
with $\Omega_0>1$ is 
\begin{eqnarray}
\tilde{D}_1(a) &=& a \times \tilde{d}_1\left({1\over a}\sqrt[3]{\Omega_0\over
\Omega_0-1}\right),\\
\tilde{d}_1(v) &=& {5\over3}v\left\{
\sqrt[4]{3}\sqrt{v^3-1}\left[{1\over3-\sqrt{3}}F(\gamma,\sin15^\circ)
- E(\gamma,\sin15^\circ)\right]+{(\sqrt{3}-1)v^2+1\over v-1+\sqrt{3}}\right\},\\
\gamma &=& {\rm Arccos}\,{v-1-\sqrt{3}\over v-1+\sqrt{3}}.
\end{eqnarray}
$v=1$ is the time at which the universe reaches maximum expansion.
As $v\rightarrow\infty$, 
$\tilde{d}_1\rightarrow1+(2/11)v^{-3}+(16/187)v^{-6}+O(v^{-9})$.

As a convenience to the reader, we collect other exact equations relevant
to low-density, flat universes from the literature (\cite{Edw72}\ 1972, 1973; 
\cite{Dab86}\ 1986, 1987; \cite{Wei72}\ 1972, 1989).  
The time-redshift relation may be written as
\begin{equation}
t(z) = {2\over3H_0}{1\over\sqrt{1-\Omega_0}}\sinh^{-1}\left(
(1+z)^{-3/2}\sqrt{1-\Omega_0\over\Omega_0}\right).
\end{equation}
The conformal time $\eta\equiv\int a^{-1}dt$ is
\begin{eqnarray}
\eta(z) &=& {1\over H_0}\int^{(1+z)^{-1}}_0{da\over a^2 H(a)}\\
&=&{1\over H_0\sqrt[4]{3}}\left[\Omega_0^2(1-\Omega_0)\right]^{-1/6}
F[\beta(z),\sin75^\circ]\\
\beta(z) &=& {\rm Arccos\,}{1+z+(1-\sqrt{3})\sqrt[3]{\Omega_0^{-1}-1}\over
1+z+(1+\sqrt{3})\sqrt[3]{\Omega_0^{-1}-1}},
\end{eqnarray}
where $H(a)$ is defined in equation (\ref{eq:H}).
The coordinate radius out to redshift $z$ is $r_1(z)=c[\eta(0)-\eta(z)]$,
where $c$ is the speed of light and the flatness of the universe has
been used.
Then the luminosity distance $d_L$ is $(1+z)r_1(z)$, the angular
diameter distance $d_A$ to an object of given physical size is 
$(1+z)^{-1}r_1(z)$, and the comoving volume per unit redshift 
per unit solid angle is
\begin{equation}
dV_c/dz\,d\Omega= {c\over H_0}{r_1^2(z)\over\sqrt{1-\Omega_0+\Omega_0(1+z)^3}}.
\end{equation}

A quantity related to the growth function is $f\equiv (a/D_1)(dD_1/da)$,
used to relate density perturbations to velocity perturbations.  
Assuming the normalization in equation (\ref{eq:norm}), one finds
(\cite{Pee84} 1984)
\begin{equation}
f(a) = {\Omega_0\over(1-\Omega_0)a^3+\Omega_0}
\left({5a\over2D_1(a)}-{3\over2}\right).
\end{equation}

In summary, the expression presented here for the growth function is
marginally more complicated than the $\Lambda=0$ case, but if one
evaluates the elliptic integrals using available numerical packages,
the exact function is quite tractable.

\noindent{\it Acknowledgements:}
I thank W.\ Hu and D.\ Spergel for useful discussions 
and acknowledge support from NSF PHY-9513835.

\clearpage
%%% \centerline{\bf Figure Captions}\bigskip
%%% \begin{figure}[h]
\begin{figure}[t]
\centerline{\epsfxsize=\textwidth\epsffile{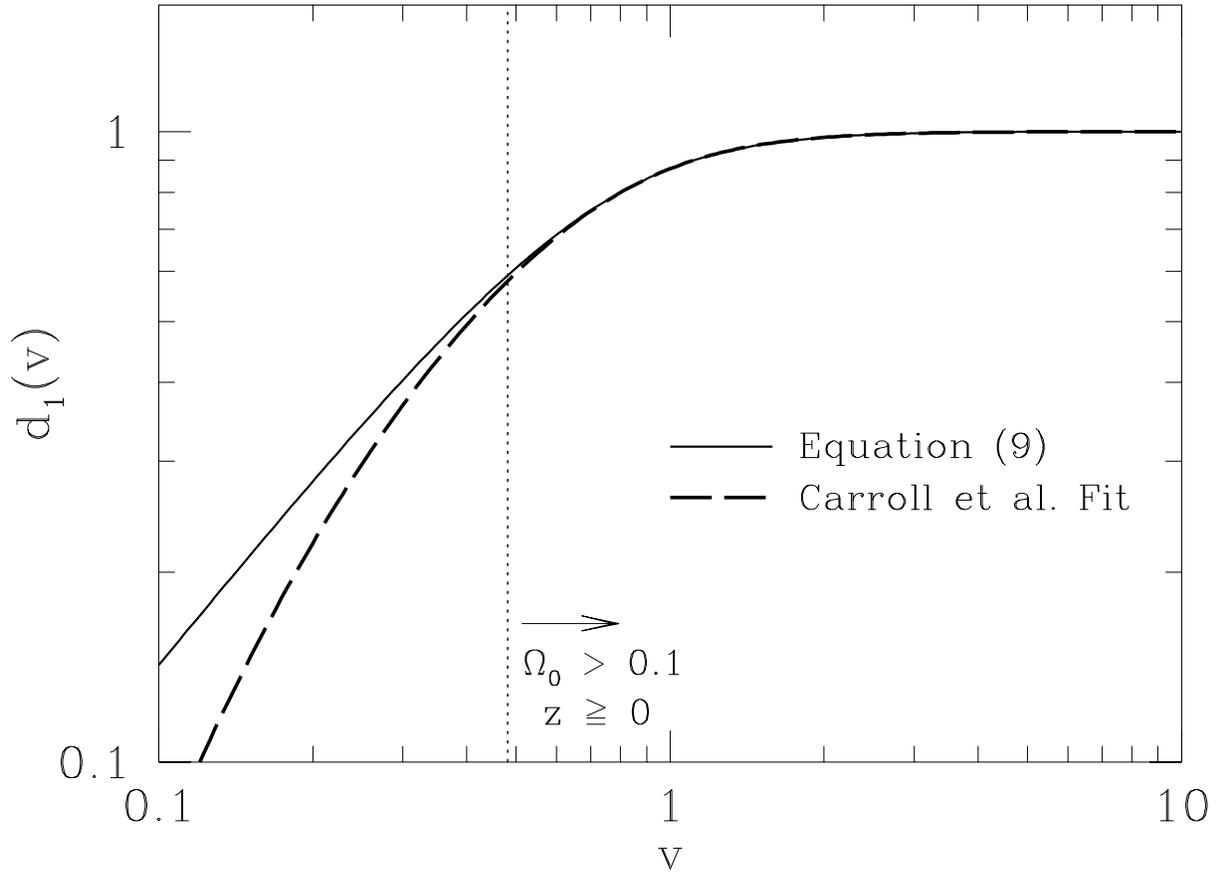}}
\caption{The function $d_1(v)$ from equation (\protect\ref{eq:done}) ({\it
solid line}).  The growth function $D_1(a)$ is
$a\,d_1(a^{-1}\protect\sqrt[3]{\Omega_0/(1-\Omega_0)}$.  The approximation of
\protect\cite{Car92}\ (1992) ({\it dashed line}), 
using $1-\Omega_0 = (v^3+1)^{-1}$, fits quite
well in the observationally relevant portion of parameter space.}
\end{figure}

\end{document}